\documentclass{article}
\usepackage{graphicx}
\newcommand{\be}{\begin{equation}}
\newcommand{\ee}{\end{equation}}
\newcommand{\bea}{\begin{eqnarray}}
\newcommand{\eea}{\end{eqnarray}}
\newcommand{\bean}{\begin{eqnarray*}}
\newcommand{\eean}{\end{eqnarray*}}
\title{Gluing 4-simplices: a derivation of the Barrett-Crane spin foam
model \\ for Euclidean quantum gravity}
\author{Daniele Oriti\thanks{d.oriti@damtp.cam.ac.uk} \\ and
\\ Ruth M. Williams\thanks{r.m.williams@damtp.cam.ac.uk} \\
 \\ Department of Applied Mathematics and
Theoretical Physics, \\ Centre for Mathematical Sciences, University of
Cambridge, \\ Wilberforce Road, Cambridge CB3 0WA, UK \\ and \\ Girton
College, Cambridge CB3 0JG, UK}

\begin{document}
\maketitle
\begin{abstract}
We derive the the Barrett-Crane spin foam model for Euclidean 4-dimensional quantum
gravity from a discretized BF theory, imposing the constraints that
reduce it to gravity at the quantum level. We obtain in this way a
precise prescription of the form of the Barrett-Crane state sum, in
the general case of an arbitrary manifold with boundary. In particular
we derive the amplitude for the edges of the spin foam from a natural
procedure of gluing different 4-simplices along a common
tetrahedron. The generalization of our results to higher dimensions is
also shown. 
\end{abstract}
\section{Introduction}

In recent years, many different approaches to the problem of finding a complete theory of 
quantum gravity have been  converging to the formalism of the
so-called spin foams  \cite{Baez,Baez2}. 
These kind of models are obtained by translating the geometric information about a (triangulated) 
manifold into the language of combinatorics and group theory, so that
the usual concepts of a 
metric and of metric properties are somehow emerging from them, and not regarded as 
fundamental. In some sense this implements in a precise way the idea of a sum over geometries, but now we are 
summing over labelled 2-complexes (spin foams), i.e. collections of faces, edges and vertices combined together 
and labelled by representations of a group (or a quantum group). A spin foam emerges when considering the evolution
 in time of spin networks \cite{ReiRov,Rov1,Rov2}, which were discovered to represent states of 
quantum general relativity at the kinematical level \cite{RovSmolin,Baez3,Ash1,Ash3,Ash4}. Spin foam models exist also for topological field
theories in different dimensions \cite{PonzReg,T-V,CrYet,CrKauYet},
and many different spin foam models have been developed for gravity 
\cite{Reis1,Reis2,Iwa1,Iwa2}. One of the most promising spin foam models for gravity in 4 
dimensions was 
proposed in \cite{BC} and is known as the Barrett-Crane state sum model. It was shown \cite{DP-F} to be related 
at the classical level to gravity, and more exactly to correspond to the Plebanski action \cite{Pleb,C-D-J}, 
which contains gravity as a sector of the solutions. This in turn can be considered as a constrained topological field 
theory, namely a BF theory \cite{BBRT}, plus a constraint on the B field.
Another result which suggests that the Barrett-Crane model is indeed related to quantum gravity 
is that the semiclassical limit of the amplitude for a 4-simplex (so in a sense the simplest possible 
manifold) gives a path integral with the action given by a form of the
Regge calculus action with the 
areas of the triangles of the triangulated manifold as variables instead of the edges of the 
triangulation \cite{BarrWill,BRW}. 
The Barrett-Crane model was originally obtained through a quantization of a 4-simplex, 
meaning a study of the correct way to translate the conditions that determine the classical 
geometry of a 4-simplex into the quantum language of representations of SO(4), which is the 
local gauge symmetry group of Euclidean gravity in 4 dimensions. In this way the quantum amplitude for a 
4-simplex was obtained and a state sum (discrete partition function) deduced from it, leaving 
some ambiguity regarding the amplitudes to be associated to the lower dimensional simplices 
(tetrahedra and triangles) in the spin foam model. In this sense the Barrett-Crane state sum  was 
more guessed than derived (for an attempt to set up a general formalism for deriving a spin foam
 model from a classical action, see \cite{FK}).

We will try to derive the Barrett-Crane model for Euclidean gravity in
 4 dimensions from a discretization of
 the SO(4) BF theory, imposing the constraints that reduce this theory to gravity 
(the Barrett-Crane constraints) at the quantum level, i.e. at the level of the representations of 
SO(4) used, and not starting from a discretization of the Plebanski action, i.e. from a constrained
 action at the classical level. 

The reasons for this approach are several: at the continuum (and classical) level the 
relation between the Plebanski action and the BF action already mentioned; at the discrete (and quantum) level, 
the fact that a complete discretization of SU(2) BF theory in 4
dimensions has been carried out 
\cite{Kawa}, and leads to the Crane-Yetter discrete topological theory, the Barrett-Crane 
model being a \lq\lq constrained doubling" of it.

Of course the best thing to do would be to discretize directly the Plebanski action, obtaining 
directly the Barrett-Crane state sum model in this way, but this is very difficult due to the non-linearity of the additional term in the B field (similar problems exist for the discretization of the 
BF theory with a cosmological constant, see \cite{O'L}), and requires further investigation. 

\section{The Barrett-Crane model} \label{sec:bc}
Let us first recall the basic elements of the Barrett-Crane work
\cite{BC}.

A geometric 4-simplex is completely and uniquely characterized (up to parallel translation and 
inversion through the origin) by a set of 10 bivectors, each corresponding to a triangle in the 
4-simplex and satisfying the following properties:
\begin{itemize}
\item the bivector changes sign if the orientation of the triangle is changed;
\item each bivector is simple, i.e. given by a wedge product of two vectors;
\item if two triangles share a common edge, the sum of the two bivectors is simple;
\item the sum (considering orientations) of the 4 bivectors corresponding to the faces of a 
tetrahedron is zero;
\item the assignment of bivectors is non-degenerate;
\item the bivectors (thought of as operators) corresponding to triangles meeting at a vertex of a 
tetrahedron satisfy the inequality $tr b_{1}[b_{2},b_{3}]\geq 0$.  

\end{itemize}

The crucial observation now is that bivectors can be thought of as being elements of the Lie 
Algebra so(4), so we can label the triangles in the triangulation with representations of so(4), 
i.e. considering the splitting $so(4)\simeq su(2) \oplus su(2)$, with pairs of spins $(j,k)$, and 
the tetrahedra in the triangulation with tensors in the product of the four spaces on its triangles. 
The point is to translate the conditions above into conditions on the representations of this 
algebra. 

The corresponding conditions on the representations were found to be the following:
\begin{itemize}
\item different orientations of a triangle correspond to dual representations;
\item the representations of the triangles are \lq\lq simple representations" of SO(4) of the form 
$(j,j)$, i.e. representations of class 1 with respect to the subgroup SO(3) \cite{V-K};
\item given two triangles, if we decompose the pair of representations into its Clebsch-Gordan 
series, the tensor for the tetrahedron is decomposed into summands which are non-zero only 
for simple representations;
\item the tensor for the tetrahedron is invariant under SO(4).   
\end{itemize}  

It was then proved \cite{Reis} that the intertwiner proposed in the original paper is unique up to 
normalization.

Out of these conditions, an amplitude for a quantum 4-simplex can be deduced and calculated 
\cite{Barr}, and it is possible to write down a spin foam model (for fixed triangulation) from 
these amplitudes:
\be Z(\Delta)\,=\,\sum_{{j_{f}}}\prod_{f}\,A_{f}\,\prod_{e}\,A_{e}\,\prod_{v}\,A_{v}^{BC} \ee
where the products are over the faces, dual to triangles, edges, dual to tetrahedra, and vertices, 
dual to 4-simplices, of the 2-complex representing the spin foam and which is dual to the 
triangulation $\Delta$ of the 4-dimensional manifold. The sum is over the spins labelling the 
triangles, and the amplitudes are the Barrett-Crane amplitude for the vertices, and suitable 
amplitudes for edges and faces of the spin foam. Since there is no complete derivation from a 
classical theory so far for this state sum, the exact amplitudes for edges and faces are not 
determined, but different models with the same Barrett-Crane amplitude for the vertices are 
proposed in \cite{DP-F-K-R} and \cite{P-R}. The problem of the choice of the amplitudes for 
the interior tetrahedra is also related to the problem of how to glue two 4-simplices along a 
common tetrahedron inside the manifold, a problem not addressed in
these works.

In this paper we try to derive the complete state sum from a constrained discretization of a 
classical theory, so that the way of gluing different 4-simplices is natural, and the corresponding 
amplitude for the edges of the spin foam is obtained automatically.  

\section{BF theory, Plebanski action, and the Barrett-Crane model}
We now review briefly the relationship between Plebanski action, BF
theory and the 
Barrett-Crane model. The so(4)-Plebanski action \cite{Pleb} (without cosmological constant) is 
given by:
\be 
S\,=\,S(\omega,B,\phi)\,=\,\int_{\mathcal{M}}\left[ B^{IJ}\,\wedge\,F_{IJ}(\omega)\,-\frac{1}{2}\phi_{IJKL}\,B^{KL}\,\wedge\,B^{IJ}\right]
\ee
where $\omega$ is an so(4)-valued connection 1-form, $\omega=\omega_{\mu}^{IJ}X_{IJ}dx^{\mu}$, $X_{IJ}$ are the generators of so(4), $F=d\omega$ is the corresponding two-form curvature, $B$ is an so(4)-valued 2-form, $B=B_{\mu\nu}^{IJ}X_{IJ}dx^{\mu}\wedge dx^{\nu}$, and $\phi_{IJKL}$ is a Lagrange multiplier. The associated equation of motion are:
\bea
\frac{\delta S}{\delta \omega}\rightarrow \mathcal{D}B\,=\,dB\,+\,[\omega,B]\,=\,0 \\
\frac{\delta S}{\delta B}\rightarrow F^{IJ}(\omega)\,=\,\phi^{IJKL}B_{KL} \\
\frac{\delta S}{\delta \phi}\rightarrow B^{IJ}\,\wedge\,B^{KL}\,=\,e\,\epsilon^{IJKL} \label{eq:constrB}
\eea
where $e=\frac{1}{4!}\epsilon_{IJKL}B^{IJ}\wedge B^{KL}$.

Thus it is evident that this theory is like a BF topological field theory, with a type of source term 
and with a non-linear constraint on the B field. In turn the relation with gravity arises because the 
constraint (~\ref{eq:constrB}) is satisfied if and only if there exists a real tetrad field 
$e^{I}=e^{I}_{\mu}dx^{\mu}$ so that one of the following equations holds:
\bea &I&\;\;\;\;\;\;\;\;\;\;B^{IJ}\,=\,\pm\,e^{I}\,\wedge\,e^{J} \\ 
&II&\;\;\;\;\;\;\;\;\;\;B^{IJ}\,=\,\pm\,\frac{1}{2}\,\epsilon^{IJ}_{KL}\,e^{K}\,\wedge\,e^{L}.   
\eea

If we restrict the field B to be always in the sector II (with the plus sign), and substitute the 
expression for B in terms of the tetrad field into the action, we obtain:
\be 
S\,=\,\int_{\mathcal{M}}\,\epsilon_{IJKL}\,e^{I}\,\wedge\,e^{J}\,\wedge\,F^{KL}
\ee
which is just the action for general relativity in the first order Palatini formalism.

The restriction on the B field is always possible classically, so the
two theories do not differ at the 
classical level, but they are different at the quantum level, since in the quantum theory one cannot
 avoid interference between different sectors. This is discussed in \cite{DP-F}.

It was shown in \cite{DP-F} that a discretization of the constraints (~\ref{eq:constrB}) which give 
gravity from BF theory prove that they are the classical 
analogue of the Barrett-Crane constraints. Consequently, we can look at the Barrett-Crane 
state sum model as a (tentative) quantization of the Plebanski action, and so strongly related 
(even if somewhat different) to gravity.   

\section{The discretized SU(2) BF theory} 
Let us now sketch the discretization of SU(2) BF theory as given in
\cite{Kawa} (Baez has pointed out some ambiguities in this discretization procedure; we refer to \cite{Baez2} and \cite{FK} for alternative approaches). 

Consider the SU(2) BF theory action, which can be thought of as being the self-dual (or anti self 
dual) part of an SO(4) BF theory action, as we will see later, 
\be S\,=\,\int_{M} B\,\wedge\,F \ee 
where $B$ is an su(2)-valued 2-form field, and $F$ is the 2-form curvature of an su(2)-valued 
connection 1-form.

Consider now a piecewise linear 4-dimensional simplicial manifold, which is given by a 
triangulation of the manifold $M$. According to the Regge calculus picture, the curvature is 
located at the different triangles t ((d-2)-dimensional simplices). Consider also the complex 
which is dual to the triangulation, having a vertex for each 4-simplex of the triangulation, an 
edge (dual link) for each tetrahedron connecting the two different 4-simplices that share it, and 
a (dual) face for each triangle in the triangulation (see Figure 1)(an earlier work using the complex dual to the triangulation is \cite{Adams}).
\begin{center} 

\includegraphics[width=8cm]{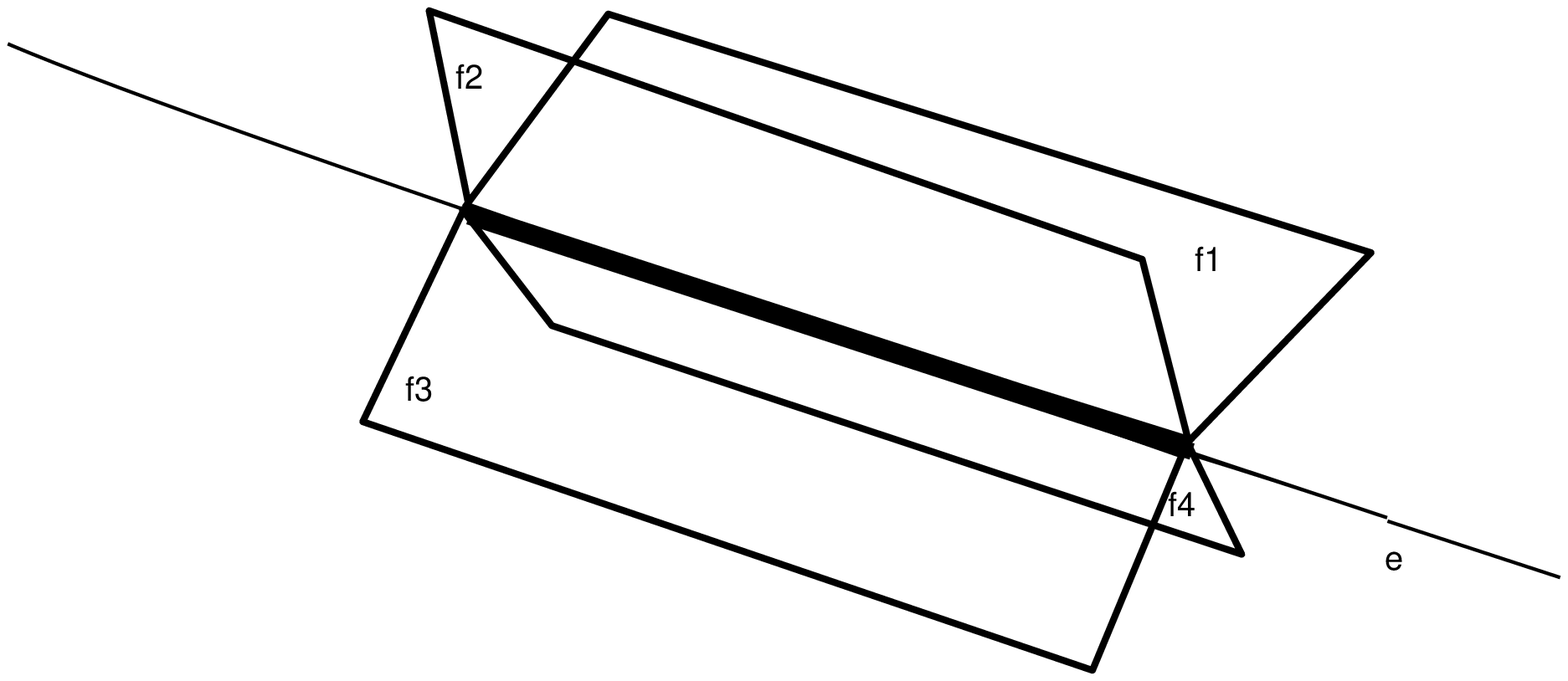}

{\normalsize Figure 1 - A dual edge e with the four dual faces meeting on it}
\end{center}

The 2-dimensional surface bounded by the 
dual links connecting the 4-simplices that share the same triangle is called a dual plaquette. It is 
easy to see that the correspondence between a triangle in the original triangulation and a dual 
plaquette is 1-1 (see Figure 2). 

\begin{center}
\includegraphics[width=5.5cm]{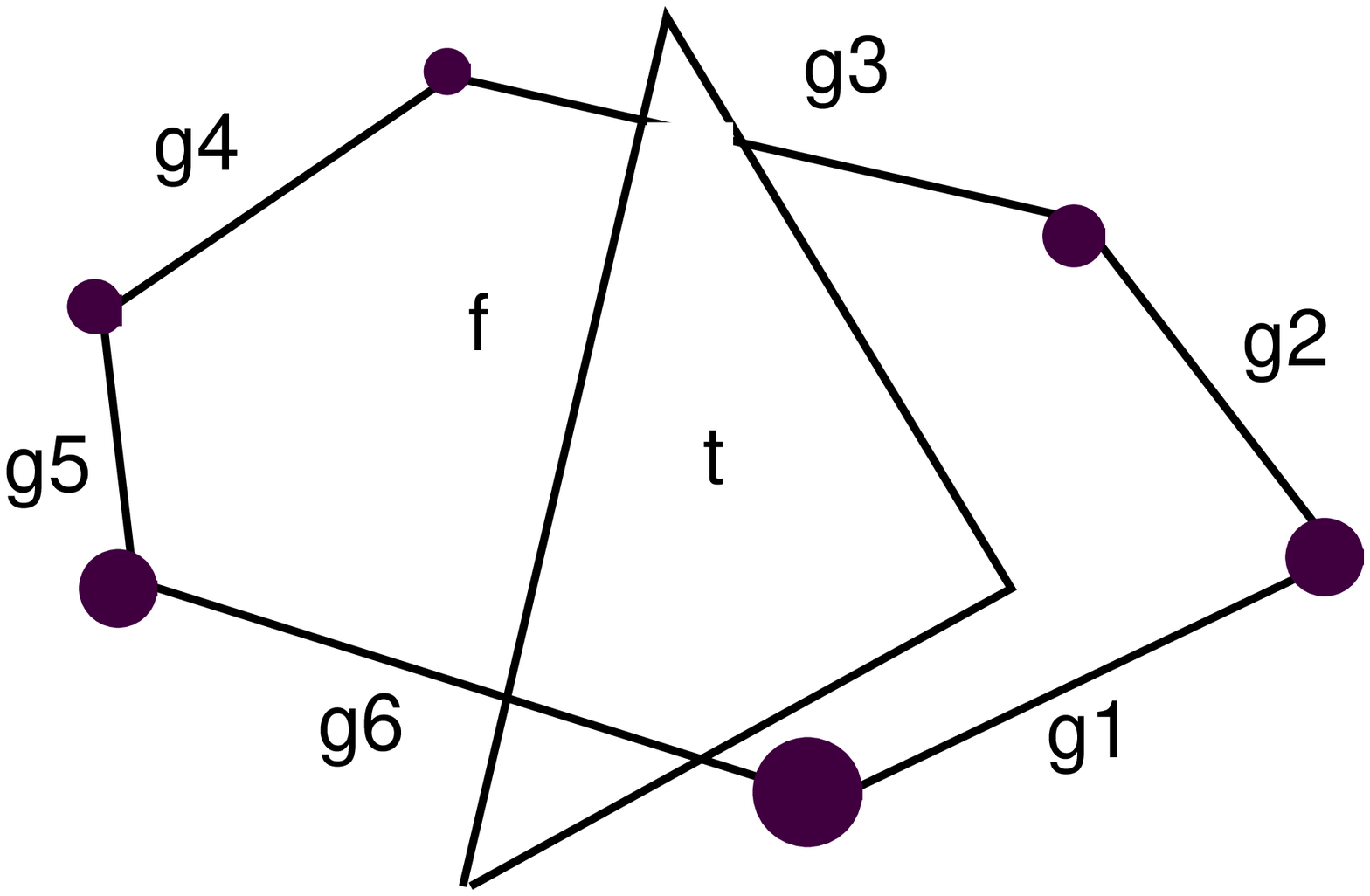}

{\normalsize Figure 2 - The dual plaquette f for the triangle t}
\end{center}

We introduce a dual link variable $U(\tilde{l})=e^{i\omega(\tilde{l})}$ for 
each dual link $\tilde{l}$. Consequently the product of dual link variables along the boundary
$\partial\tilde{P}$ of a dual plaquette $\tilde{P}$ leads to a curvature located at the center of  
the dual plaquette, i.e. at the center of the triangle t. 

We define the curvature $F(t)$ located on the triangle t by the equation:
\be
\prod_{\tilde{l}\in \partial\tilde{P}} U(\tilde{l}) \equiv e^{i\,F(t)}.
\ee

We then approximate the 2-form field $B$ with a distributional field $B(t)$ with values on the 
triangles of the original triangulation.
Note that this gives an exact theory for a topological field theory like the BF one, but it 
represents only an approximation for a non-topological theory like gravity. Nevertheless this 
approximation would be better and better when we refine the
triangulation, or sum over all the possible different triangulations,
which would be the next step after constructing a spin foam model for
a given triangulation.

The discretized action for BF theory is then  
\be 
S\,=\,\frac{1}{2}\sum_{t} B^{I} F^{I}\,=\,\sum_{t} 
tr 
\left(
-i\,B(t) \left[ \ln \prod_{\tilde{l}\in \tilde{P}} U(\tilde{l}) \right]\right)
\ee
where now the indices $I$ refer to su(2) algebra values.
 
We then impose the following constraint:
\be
\left[\left(\prod_{\tilde{l}\in \tilde{P}} U(\tilde{l})\right)\,B\,\left(\prod_{\tilde{l}\in \tilde{P}} U(\tilde{l})\right)^{\dagger}\right]^{\alpha\beta}\,=\,B^{\alpha\beta}
\ee
which is equivalent to imposing on the discrete partition function the BF equation of motion on 
the $F$ field which says that the holonomy of the curvature vanishes. 

This constraint is equivalent to the relation:
\be [F,B]\,=\,i\,\epsilon_{IJK}F^{I}\,B^{J}\,\frac{\sigma^{K}}{2}\,=\,0 \ee
or $B^{I}\propto F^{I}$.
Taking into account the parallel and antiparallel nature of 
$B^{I}$ and $F^{I}$ this constraint can be rewritten as 
\be
\frac{B^{3}}{\mid B \mid}\left[
\prod_{I=1}^{2}\,\delta\left(\frac{F^{I}}{\mid F \mid}\,
+\,\frac{B^{I}}{\mid B \mid}\right)\,+\,\prod_{I=1}^{2}\,\delta\left(\frac{F^{I}}{\mid F \mid}\,-\,\frac{B^{I}}{\mid B \mid}\right)\right]
\ee
where the term $\frac{B^{3}}{\mid B \mid}$ is needed to keep rotational invariance of the 
expression.

The necessity of another kind of constraint is clear from the following argument.
Consider the identity 
\be
e^{i4\pi n  L}\,=\,1
\ee
where $L=\frac{F^{I}}{\mid F\mid}\frac{\sigma^{I}}{2}$.

Inserting this into the expression for the action, and using the fact
that $F$ and $B$ are parallel 
leads to 
\bea
S\,=\,\sum_{t}\,tr\left(-i\,B(t)\,\ln\,e^{iF(t)}\right)=\sum_{t}\,tr\left(-i\,B(t)\,\ln\,e^{iF(t)+i4\pi n I}\right) \nonumber \\ 
=\,S\,+\,\frac{1}{2}\,\sum_{t}4\pi n \mid B(t)\mid. \eea
Thus, imposing the single valuedness of $e^{iS}$ (and hence of the partition function) we have an
 additional constraint for the B field to be of integer absolute value $N=2J$, with 
$J$ half-integer.

Finally we can write down the partition function for the SU(2) lattice theory with the above 
constraints as:
\be
Z\,=\,\int\,\mathcal{D}U\mathcal{D}B\,\delta\left(\left(\prod_{\tilde{l}\in \tilde{P}} U(\tilde{l})\right) B \left( \prod_{\tilde{l}\in \tilde{P}} U(\tilde{l})\right)^{\dagger}\,-\,B\right)\sum_{N}\delta(\mid B\mid\,-\,N) \, e^{iS}. \ee 

It is possible to prove \cite{Kawa} that this partition function is invariant under gauge 
transformation on the lattice.

Evaluating the B integral, we obtain:
\be
Z\,=\,\int\mathcal{D}U\sum_{J}8J\cos(J\mid F\mid).
\ee
Using the known formula for the character $\chi_{J}$ of the spin-J representation of SU(2) 
\be
\chi_{J}(e^{iF^{I}J_{I}})\,=\,\frac{\sin\left((2J+1)\frac{\mid F\mid}{2}\right)}{\sin\frac{\mid F\mid}{2}},
\ee
we can recast it in the form:
\be 
Z\,=\,\int\mathcal{D}U \, \prod_{t} \sum_{J} (2J+1)\,\chi_{J}\left(\prod_{\tilde{l}\in \tilde{P}} U(\tilde{l})\right). \ee
This expression is just formal because the summation is not convergent, but can be easily 
regularized. We will discuss the regularization issue later. 

\section{The discretized SO(4) BF theory} \label{sec:BF}
Let us now turn to the case of the SO(4) BF theory.

It is well known that the double covering of the SO(4) algebra, the Spin(4) algebra, is isomorphic to a direct
product of two SU(2) algebras
\be Spin(4)\simeq SU(2)_{L}\times SU(2)_{R}. \ee  Since we are interested in the connection with gravity, and so in
only some representations of this group, the simple representations, we can use this decomposition also in our case,
 and so work with the Spin(4) group, because at the end the imposition of the constraints will give us the same result as 
if we had started from an SO(4) theory, the reason being that the set of simple representations of SO(4) coincides with the set of simple representations of Spin(4).

 Thus we can split the Spin(4) BF theory action into a sum of the
 SU(2) chiral parts:
\be S\,=\,\int_{\mathcal{M}} B_{IJ}F^{IJ}\,=\,\int_{\mathcal{M}}  B_{I}^{+}F_{I}^{+}\,+\,\int_{\mathcal{M}} 
 B_{I}^{-}F_{I}^{-}. \ee
Consequently the Spin(4) BF partition function gets factorized into a
product of two SU(2) partition functions:
\be Z(Spin(4))\,=\,Z(SU(2))_{L}\,Z(SU(2))_{R} \ee
and at the discretized level we can write (dropping the ``L'' and ``R'' subscripts):
\bea 
\lefteqn{Z(Spin(4))\,=\,Z(SU(2))\,Z(SU(2))} \nonumber \\ &=&\,\int\mathcal{D}U \,
\prod_{t} \sum_{j} (2j+1)\,\chi_{j}\left(\prod_{\tilde{l}\in
\tilde{P}} U(\tilde{l})\right)\,\int\mathcal{D}U' \, \prod_{t} \sum_{k}
(2k+1)\,\chi_{k}\left(\prod_{\tilde{l}\in \tilde{P}}
U'(\tilde{l})\right) \nonumber \\ &=& \int\mathcal{D}U\mathcal{D}U' \, \prod_{t}
\sum_{j,k} (2j+1)(2k+1)\,\chi_{j}\left(\prod_{\tilde{l}\in \tilde{P}}
U(\tilde{l})\right)\chi_{k}\left(\prod_{\tilde{l}\in \tilde{P}}
U'(\tilde{l})\right),
\eea
so  we are assigning two independent SU(2) variables to each dual link.

Now the product of characters of two representations $j$ and $k$ is
given by the character of the direct product representation $j\times
k$ of the group $SU(2)\times SU(2)$:
\be 
\chi_{j\times k}(\prod (U,U'))\,=\,\chi_{j}(\prod U)\,\chi_{k}(\prod
U').
\ee

Thus we have:
\be
Z(Spin(4))\,=\,\int dU\,dU'\,\prod_{t}\,\sum_{j,k}
(2j+1)(2k+1)\,\chi_{j\times k}(\prod (U,U')). \ee

Now we note that the double integral over SU(2) is equivalent,
because of the isomorphism mentioned, to an integral over Spin(4) and
the sum above is:
\be
\sum_{j,k}
(2j+1)(2k+1)\,\chi_{j\times k}(\prod
(U,U'))\,=\,\sum_{J}\,dim_{J}\,\chi_{J}(\prod g) \ee
where $J$ is the highest weight of the general $(j,k)$ representation of Spin(4) \cite{V-K},
and the assignment of the pair of SU(2) elements $(U,U')$ is
equivalent to an assignment of a Spin(4) group element $g$.

In the end, we have the following expression for the discretized
partition function of Spin(4) BF theory:
\be
Z_{BF}(Spin(4))\,=\,\int_{Spin(4)}dg\,\prod_{\sigma}\sum_{J_{\sigma}}\,dim_{J_{\sigma}}\,\chi_{J_{\sigma}}(\prod_{e}
g_{e}) \ee
where the first product is over the plaquettes in the dual complex (remember the 1-1 correspondence between triangles and plaquettes), the sum is over (the highest weight of) the 
representations of Spin(4), and the last product is over the edges of the dual complex to which the group 
element is assigned.  

The partition function for the SO(4) BF theory is consequently obtained considering only the representation for which the components of the vectors $J_{\sigma}$ are all integers. 

\section{Constraining of the BF theory and the Barrett-Crane model for a single 4-simplex} \label{sec:1BC}
Before going on we clarify what is exactly the location of the $g$
variables; consider a 4-simplex;
it has 5 different 3-simplices (tetrahedra) in it, (1-2-3-4), (4-5-6-7), (7-3-8-9), (9-6-2-10), (1-5-8-10) (the numbers 
label the triangles in the tetrahedra of the 4-simplex), each of which is
given by 4 2-simplices (triangles), and each d-simplex is glued to another one along a common (d-1)-simplex. Thus a generic 4-simplex has 5
tetrahedra and 10 triangles in it (see Figure 3).
 Each dual link goes from a
4-simplex to a neighbouring one through the shared tetrahedron, so we
have 5 dual links coming out from a 4-simplex.

\begin{center}

\includegraphics[width=9cm]{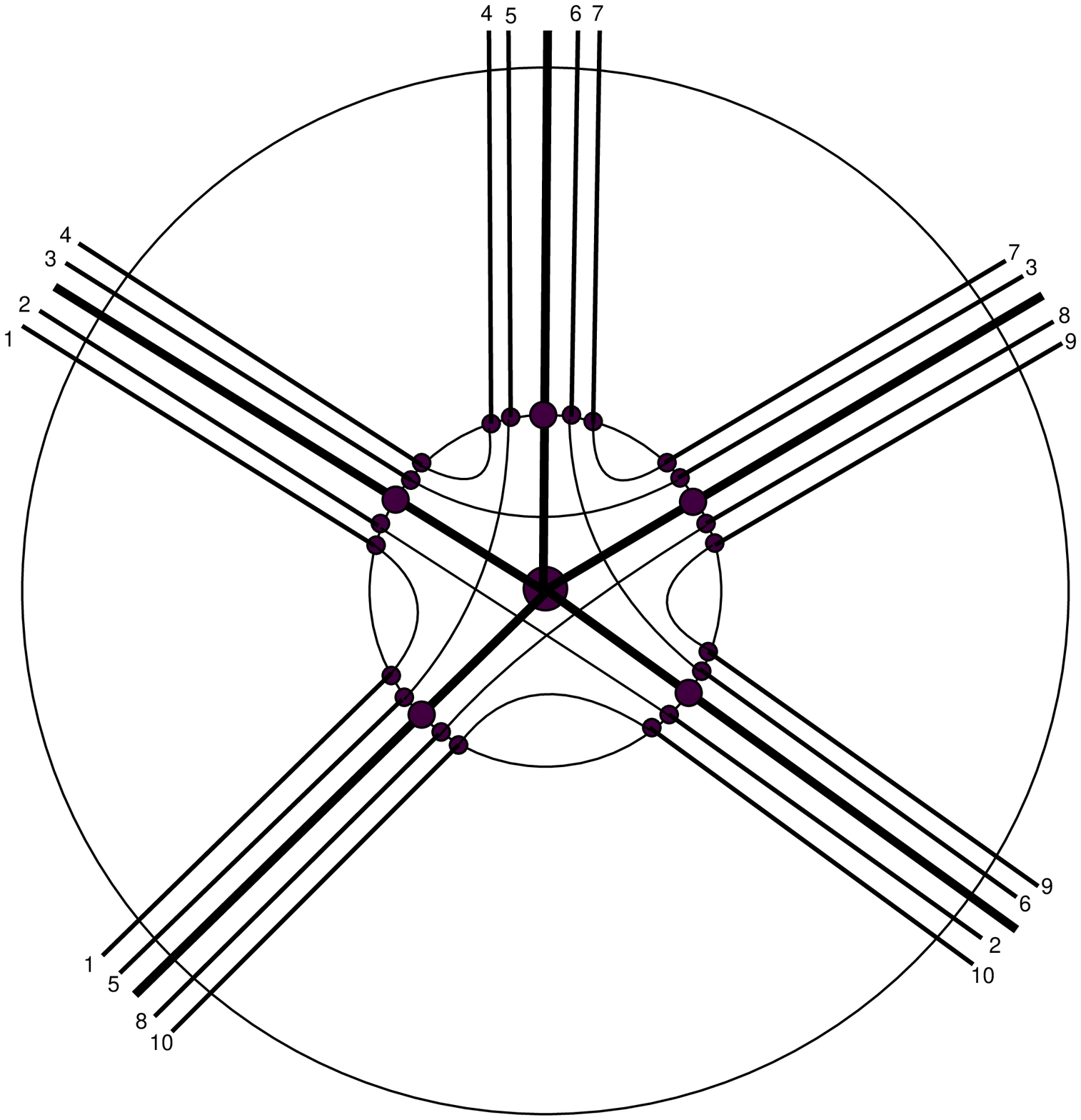}

{\normalsize Figure 3 - Schematic representation of a 4-simplex; the thick lines represent the 5 tetrahedra and the thin lines the triangles}
\end{center}

 We can assign two dual
link variables to each dual link dividing it into two segments going
from the center of each 4-simplex to the center of the boundary
tetrahedron, i.e. we assign one group element $g$ to each of them (see
Figure 4).

\begin{center}

\includegraphics[width=7cm]{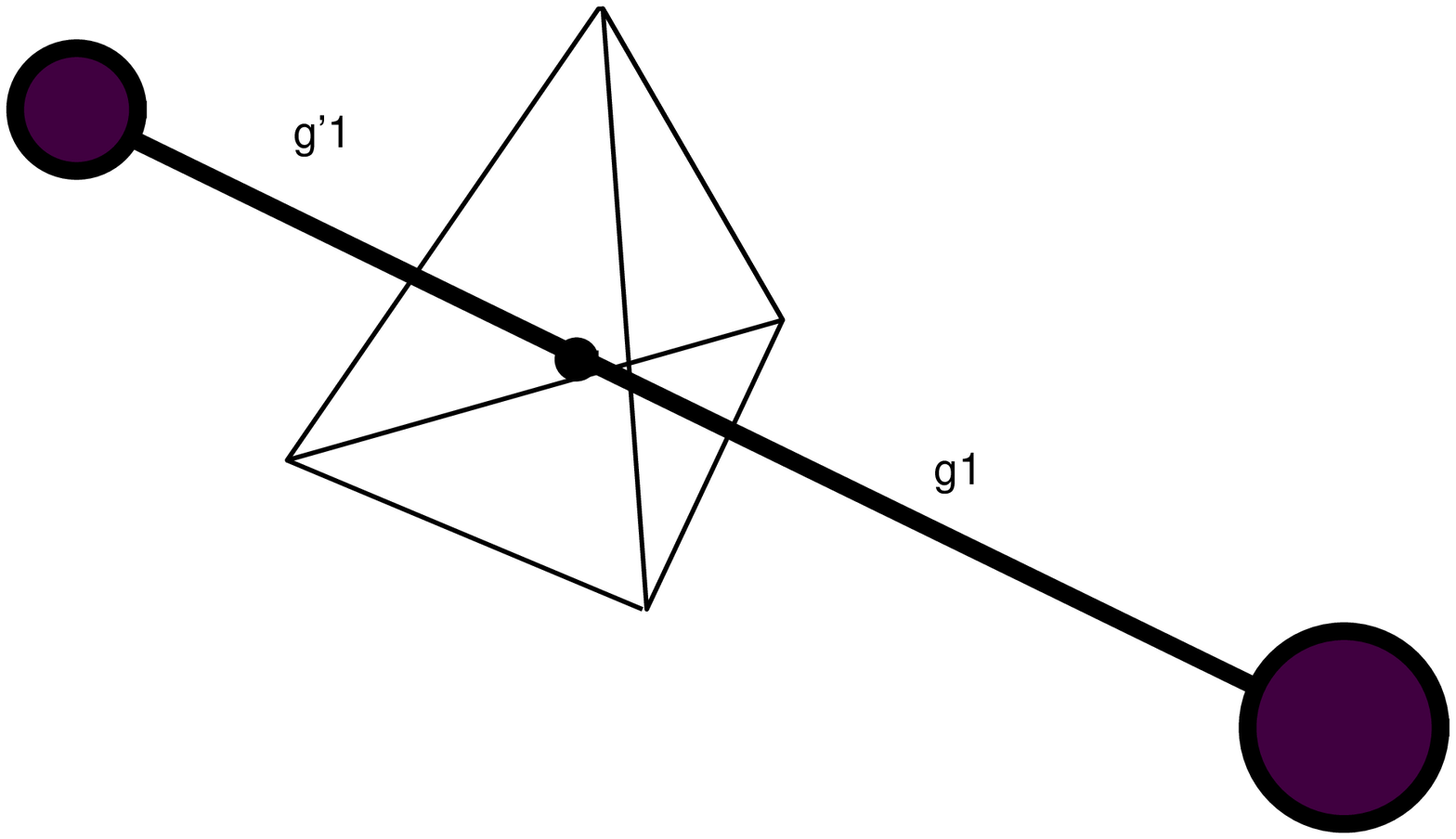} 

{\normalsize Figure 4 - The dual link corresponding to the tetrahedron on which 2 4-simplices meet}
\end{center}
Consider now a dual plaquette. It is given by a number, say, m of dual
links each divided into two segments, so there are 2m dual link
variables on the boundary of each plaquette. When a tetrahedron sharing the triangle to which the plaquette corresponds is on the boundary of the manifold, the plaquette results in being truncated by the boundary, and there will be edges exposed on it (not connecting 4-simplices). To each of these exposed edges we also assign a group variable.

We now make use of the character decomposition formula which
decomposes the character of a given representation of a product of
group elements into a product of (Wigner) D-functions in that representation:
\be 
\chi_{J_{\sigma}}\left(\prod_{\tilde{l}\in \partial\tilde{P}}
g_{e}(\tilde{l})\right)\,=\,\sum_{\{k\}}\prod_{i}D_{k_{i}k_{i+1}}^{J_{\sigma}}\left(g_{e_{i}}\right)\;\;\;{\it
with}\;\;\;k_{1}=k_{2m+1}
\ee
where the product on the $i$ index goes around the boundary of the
dual plaquette surrounding the triangle labelled by $J_{\sigma}$, and
there is a D-function for each group element assigned to a dual link and to the edges exposed on the boundary .

We choose real representations of Spin(4) (this is always possible).
Note that this can be seen as a way to implement automatically the first of the Barrett-Crane constraints, so that there 
will be no need to impose it explicitly in the following. Thus we have:
\be
Z_{BF}(Spin(4))\,=\,\left(\prod_{e}\int_{Spin(4)}dg_{e}\right)\,\prod_{\sigma}\sum_{J_{\sigma},\{k\}}\,dim_{J_{\sigma}}\,\prod_{i}D_{k_{i}k_{i+1}}^{J_{\sigma}}\left(g_{e_{i}}\right).
\ee
Consider now a single 4 simplex. Note that in this case all the tetrahedra are on the boundary of the manifold, which is
 given by the interior of the 4-simplex.
Writing down explicitly all the products of D-functions and labelling
 the indices appropriately, we can write down the partition function for the
Spin(4) BF theory on a manifold consisting of a single 4-simplex in the
following way:
\bea 
\lefteqn{Z_{BF}(Spin(4))=} \nonumber \\ &=&\sum_{\{J_{\sigma}\},\{k_{e}\}}\left(\prod_{\sigma}dim_{J_{\sigma}}\right)\prod_{e}\int_{Spin(4)}dg_{e}\,D_{k_{e1}m_{e1}}^{J_{1}^{e}}D_{k_{e2}m_{e2}}^{J_{2}^{e}}D_{k_{e3}m_{e3}}^{J_{3}^{e}}D_{k_{e4}m_{e4}}^{J_{4}^{e}}\;\left(\prod_{\tilde{e}}D_{il}^{J}\right)\;\;\;\;. 
\eea
The situation is now as follows: we have a contribution for each of the 5 edges of the dual complex, corresponding to 
the tetrahedra of the triangulation, each of them made of a product of the 4 D-functions for the 4 representations 
labelling the 4 faces incident to an edge, corresponding to the 4 triangles of the tetrahedron. There is an extra product 
over the faces with a weight given by the dimension of the representation labelling that face, and the indices of the 
Wigner D-functions refer one to the center of the 4-simplex, one end of the dual edge, and the other to a tetrahedron 
on the boundary, the other end of the dual edge. There is also an additional product of D-functions, one for each group element assigned to an edge exposed on the boundary, and not integrated over because we are working with fixed connection on the boundary.  

Now we want to go from BF theory to gravity (Plebanski) theory by
imposing the Barrett-Crane constraints on the BF partition
function. These are quantum constraints on the representations of SO(4) which
are assigned to each triangle of the triangulation, so they can be
imposed at this \lq\lq quantum" level.
The constraints are essentially two: the simplicity constraints,
saying that the representations by which we label the triangles are to
be chosen from the simple representations of SO(4) (Spin(4)), and the closure
constraint, saying that the tensor assigned to each tetrahedron has to
be an invariant tensor of SO(4) (Spin(4)). As we have chosen real representations, there is no need to impose the 
first constraint of section ~\ref{sec:bc}, and the third one will be imposed automatically in the following. We can 
implement the second constraint
at this level by requiring that all the representation functions have
to be invariant under the subgroup SO(3) of SO(4), so realizing these representations in the space of harmonic 
functions over the coset ${SO(4)}/{SO(3)}\simeq S^{3}$, which was proven in \cite{F-K-P}\cite{F-K} to be a 
complete characterization of the simple representations of SO(D) for any dimension D. We then implement the fourth 
constraint
by requiring that the amplitude for a tetrahedron is invariant under a
general SO(4) transformation.
We note that these constraints have the effect of breaking the topological invariance of the theory.
Moreover, from now on we can replace the integrals over Spin(4) with integrals over SO(4), and the sum with a sum 
over the SO(4) representations only. 

Consequently we write:
\bea 
\lefteqn{Z_{BC}=\sum_{J_{\sigma},\{k_{e}\}}\left(\prod_{\sigma}dim_{J_{\sigma}}\right)}
\nonumber \\ && \prod_{e}\int_{SO(4)}dg_{e}\int_{SO(3)}dh_{1}\int_{SO(3)}dh_{2}\int_{SO(3)}dh_{3}\int_{SO(3)}dh_{4}\int_{SO(4)}dg'_{e}
\nonumber \\ && D_{k_{e1}m_{e1}}^{J_{1}^{e}}(g_{e}h_{1}g'_{e})D_{k_{e2}m_{e2}}^{J_{2}^{e}}(g_{e}h_{2}g'_{e})D_{k_{e3}m_{e3}}^{J_{3}^{e}}(g_{e}h_{3}g'_{e})D_{k_{e4}m_{e4}}^{J_{4}^{e}}(g_{e}h_{4}g'_{e}) \left(\prod_{\tilde{e}}D\right)\;\;\;\;\;\;\;\;\nonumber \\ 
&=& \sum_{J_{\sigma},\{k_{e}\}}\left(\prod_{\sigma}dim_{J_{\sigma}}\right)\prod_{e}\,A_{e}\left(\prod_{\tilde{e}}D\right).  
\eea
Let us consider now the amplitude for each edge $e$ of the dual
complex:
\bea
\lefteqn{A_{e}=\int_{SO(4)}dg_{e}\int_{SO(3)}dh_{1}\int_{SO(3)}dh_{2}\int_{SO(3)}dh_{3}\int_{SO(3)}dh_{4}\int_{SO(4)}dg'_{e}\,}
\nonumber \\ 
&& D_{k_{e1}m_{e1}}^{J_{1}^{e}}(g_{e}h_{1}g'_{e})D_{k_{e2}m_{e2}}^{J_{2}^{e}}(g_{e}h_{2}g'_{e})D_{k_{e3}m_{e3}}^{J_{3}^{e}}(g_{e}h_{3}g'_{e})D_{k_{e4}m_{e4}}^{J_{4}^{e}}(g_{e}h_{4}g'_{e})
\eea
for a particular tetrahedron (edge) made out of the triangles 1,2,3,4, say, and write  the integrals using the decomposition property for the
representation function of a product of group elements:
\bea
\lefteqn{A_{e}=\int_{SO(4)}dg_{1}D_{k_{1}l_{1}}^{J_{1}}(g_{1})D_{k_{2}l_{2}}^{J_{2}}(g_{1})D_{k_{3}l_{3}}^{J_{3}}(g_{1})D_{k_{4}l_{4}}^{J_{4}}(g_{1})} \nonumber \\ &\times&\int_{h_{1}}dh_{1}D_{l_{1}i_{1}}^{J_{1}}(h_{1})\int_{SO(3)}dh_{2}D_{l_{2}i_{2}}^{J_{2}}(h_{2})\int_{SO(3)}dh_{3}D_{l_{3}i_{3}}^{J_{3}}(h_{3})\int_{SO(3)}dh_{4}D_{l_{4}i_{4}}^{J_{4}}(h_{4})
\nonumber \\ &\times&\int_{SO(4)}dg'_{1}\,D_{i_{1}m_{1}}^{J_{1}}(g'_{1})D_{i_{2}m_{2}}^{J_{2}}(g'_{1})D_{i_{3}m_{3}}^{J_{3}}(g'_{1})D_{i_{4}m_{4}}^{J_{4}}(g'_{1})\eea
where the sum over repeated indices is understood. We have now to perform the different integrals. 

The integral of a product of 4 D-functions is given by:
\be
\int_{SO(4)}\,dg\,D_{\alpha_{1}\beta_{1}}^{j_{1}}(g)D_{\alpha_{2}\beta_{2}}^{j_{2}}(g)D_{\alpha_{3}\beta_{3}}^{j_{3}}(g)D_{\alpha_{4}\beta_{4}}^{j_{4}}(g)\,=\,\sum_{J}\,C^{j_{1}j_{2}j_{3}j_{4}J}_{\alpha_{1}\alpha_{2}\alpha_{3}\alpha_{4}}\,C^{j_{1}j_{2}j_{3}j_{4}J}_{\beta_{1}\beta_{2}\beta_{3}\beta_{4}} \label{eq:C} \ee
where the $C$ functions for all $J$'s are an orthonormal basis for the space of the SO(4) invariant tensors that are 
intertwiners between the 4 different representations $j_{1}$, $j_{2}$,
$j_{3}$, $j_{4}$, so that (~\ref{eq:C}). They are given by:
\be
C^{j_{1}j_{2}j_{3}j_{4}J}_{\alpha_{1}\alpha_{2}\alpha_{3}\alpha_{4}}\,=\,\sqrt{dim_{J}}\,C^{j_{1}j_{2}J}_{\alpha_{1}\alpha_{2}\alpha}\,C^{j_{3}j_{4}J}_{\alpha_{3}\alpha_{4}\alpha}
\ee
where the $C_{\alpha_{i}\alpha_{k}}^{j_{i}j_{k}J}$ are Wigner 3-j
symbols for SO(4), normalized so that
 
$C_{\alpha_{i}\alpha_{k}\alpha}^{j_{i}j_{k}J}C_{\alpha_{i}\alpha_{k}\beta}^{j_{i}j_{k}K}=\delta_{JK}\delta_{\alpha\beta}$.

The integral over the subgroup SO(3) of a representation function of a subgroup element in a representation J of the 
group SO(4) is given by \cite{V-K}:
\be
\int_{SO(3)}dh\,D_{\alpha\beta}^{J}(h)\,=\,w_{\alpha}^{J}w_{\beta}^{J}
\ee
where $w_{\alpha}^{J}$ is a normalized SO(3) invariant vector in 4 dimensions in the irreducible representation J of 
SO(4). Since such a vector exists (is non vanishing) only if the representation J is simple, the effect of the integrations over 
SO(3) is to project the intertwiners $C$ into the one-dimensional vector space of intertwiners between simple 
representations of SO(4). 

Consequently we obtain:
\bea 
A_{e}=\sum_{I,L}C_{k_{1}k_{2}k_{3}k_{4}}^{J_{1}J_{2}J_{3}J_{4}I}C_{l_{1}l_{2}l_{3}l_{4}}^{J_{1}J_{2}J_{3}J_{4}I}w^{J_{1}}_{l_{1}}w^{J_{2}}_{l_{2}}w^{J_{3}}_{l_{3}}w^{J_{4}}_{l_{4}}w^{J_{1}}_{i_{1}}w^{J_{2}}_{i_{2}}w^{J_{3}}_{i_{3}}w^{J_{4}}_{i_{4}}
\nonumber \\ C_{i_{1}i_{2}i_{3}i_{4}}^{J_{1}J_{2}J_{3}J_{4}L}C_{m_{1}m_{2}m_{3}m_{4}}^{J_{1}J_{2}J_{3}J_{4}L}.
\eea

As we said, the projection of the intertwiner
$C_{l_{1}l_{2}l_{3}l_{4}}^{J_{1}J_{2}J_{3}J_{4}I}w^{J_{1}}_{l_{1}}w^{J_{2}}_{l_{2}}w^{J_{3}}_{l_{3}}w^{J_{4}}_{l_{4}}$
vanishes unless all the J's and the I (or the L) are simple. When this
happens, the result is given by \cite{V-K}:
\be 
C_{\alpha_{1}\alpha_{2}\alpha_{3}\alpha_{4}}^{J_{1}J_{2}J_{3}J_{4}I}w^{J_{1}}_{\alpha_{1}}w_{\alpha_{2}}^{J_{2}}w^{J_{3}}_{\alpha_{3}}w_{\alpha_{4}}^{J_{4}}=\frac{1}{\sqrt{\Delta_{J_{1}}\Delta_{J_{2}}\Delta_{J_{3}}\Delta_{J_{4}}}}
\ee
where $\Delta_{j}=dim_{j}$, so the amplitude for a single tetrahedron on the boundary is:
\bea 
A_{e}\,=\,\sum_{simple\;I,L}\frac{1}{\Delta_{J_{1}}\Delta_{J_{2}}\Delta_{J_{3}}\Delta_{J_{4}}}C_{k_{1}k_{2}k_{3}k_{4}}^{J_{1}J_{2}J_{3}J_{4}I}C_{m_{1}m_{2}m_{3}m_{4}}^{J_{1}J_{2}J_{3}J_{4}L}
\nonumber
\\
=\frac{1}{\Delta_{J_{1}}\Delta_{J_{2}}\Delta_{J_{3}}\Delta_{J_{4}}}B_{k_{1}k_{2}k_{3}k_{4}}^{J_{1}J_{2}J_{3}J_{4}}B_{m_{1}m_{2}m_{3}m_{4}}^{J_{1}J_{2}J_{3}J_{4}},
\eea
where the $B$'s are the Barrett-Crane intertwiners, defined in \cite{BC}, and shown to be unique up to scaling in 
\cite{Reis}, and from now on the sums are over simple representations only.

Note that the simplicity of the representations labelling the
tetrahedra (the third of the Barrett-Crane constraints) comes 
automatically, without the need to impose it explicitly.
 
We note also that because of the projection above and the consequent restriction to the simple representations of the 
group, the result we end with is independent of having started from the Spin(4) or the SO(4) BF partition 
function, as we anticipated.

We see that each tetrahedron on the boundary of the 4-simplex contributes with two Barrett-Crane
intertwiners, one with indices referring to the centre of the 4-simplex
and the other indices referring to the centre of the tetrahedron
itself (see Figure 5).
\begin{center}

\includegraphics[width=7.5cm]{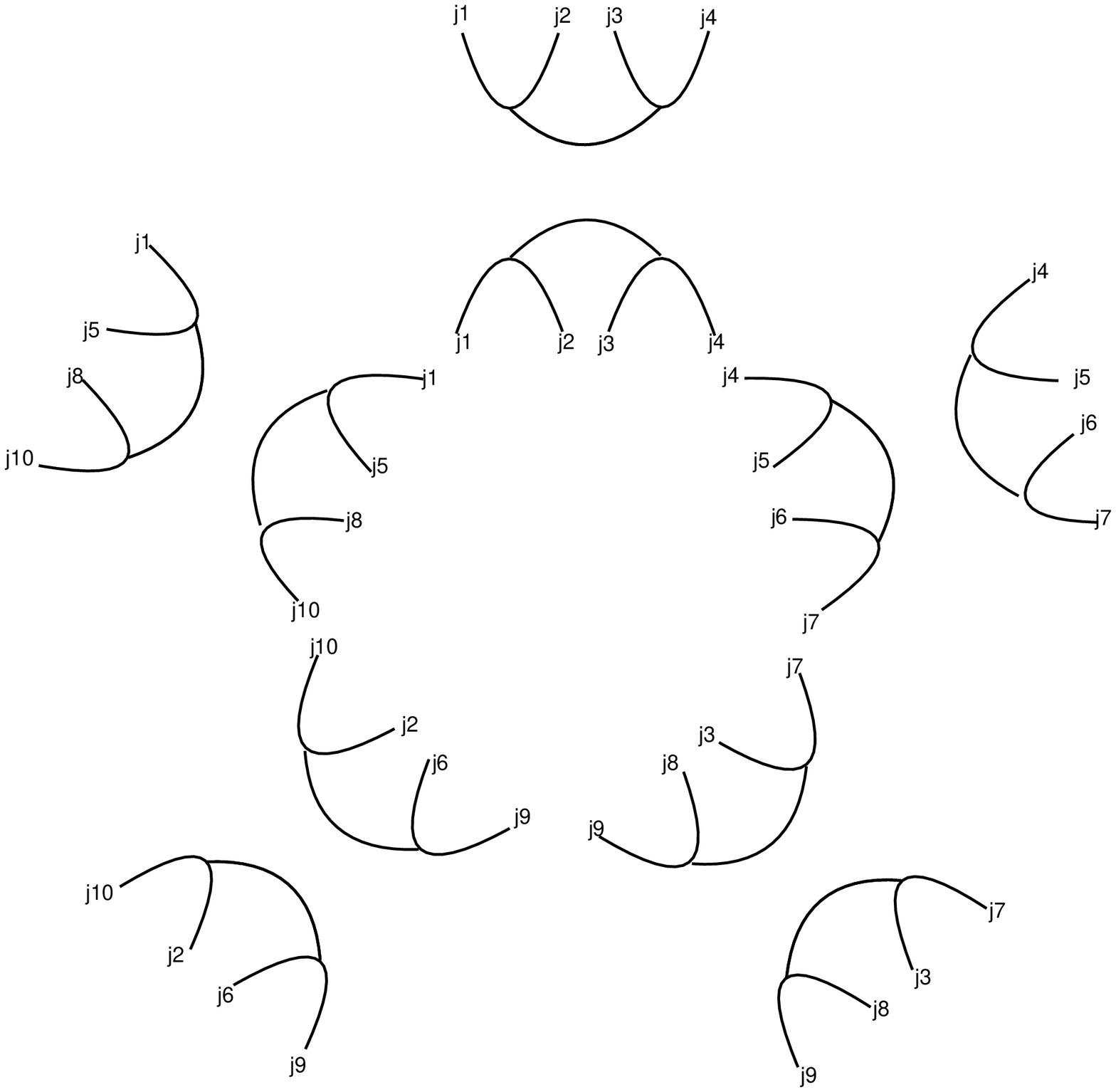}

{\normalsize Figure 5 - Diagram of a 4-simplex, indicating the two Barrett-Crane intertwiners assigned to each tetrahedron}
\end{center}
The partition function for this theory (taking into account all the
different tetrahedra) is then given by:
\bea 
\lefteqn{Z_{BC}=\sum_{\{J\},\{k\},\{n\},\{l\},\{i\},\{m\}}\Delta_{J_{1}}...\Delta_{J_{10}}\frac{1}{(\Delta_{J_{1}}...\Delta_{J_{10}})^{2}}}
\nonumber
\\
&&
B_{k_{1}k_{2}k_{3}k_{4}}^{J_{1}J_{2}J_{3}J_{4}}B_{l_{4}l_{5}l_{6}l_{7}}^{J_{4}J_{5}J_{6}J_{7}}B_{n_{7}n_{3}n_{8}n_{9}}^{J_{7}J_{3}J_{8}J_{9}}B_{h_{9}h_{6}h_{2}h_{10}}^{J_{9}J_{6}J_{2}J_{10}}B_{i_{10}i_{8}i_{5}i_{1}}^{J_{10}J_{8}J_{5}J_{1}} \nonumber
\\ && B_{m_{1}m_{2}m_{3}m_{4}}^{J_{1}J_{2}J_{3}J_{4}}B_{m_{4}m_{5}m_{6}m_{7}}^{J_{4}J_{5}J_{6}J_{7}}B_{m_{7}m_{3}m_{8}m_{9}}^{J_{7}J_{3}J_{8}J_{9}}B_{m_{9}m_{6}m_{2}m_{10}}^{J_{9}J_{6}J_{2}J_{10}}B_{m_{10}m_{8}m_{5}m_{1}}^{J_{10}J_{8}J_{5}J_{1}}\left(\prod_{\tilde{e}}D\right).
\eea

Now the product of the five Barrett-Crane intertwiners with indices
$m$ gives just the Barrett-Crane amplitude for the 4-simplex which the
indices refer to, given by a 15j-symbol constructed out of the 10 labels of the triangles and the 5 labels of the 
tetrahedra (see Figure 6), 
\begin{center}
\includegraphics[width=7.5cm]{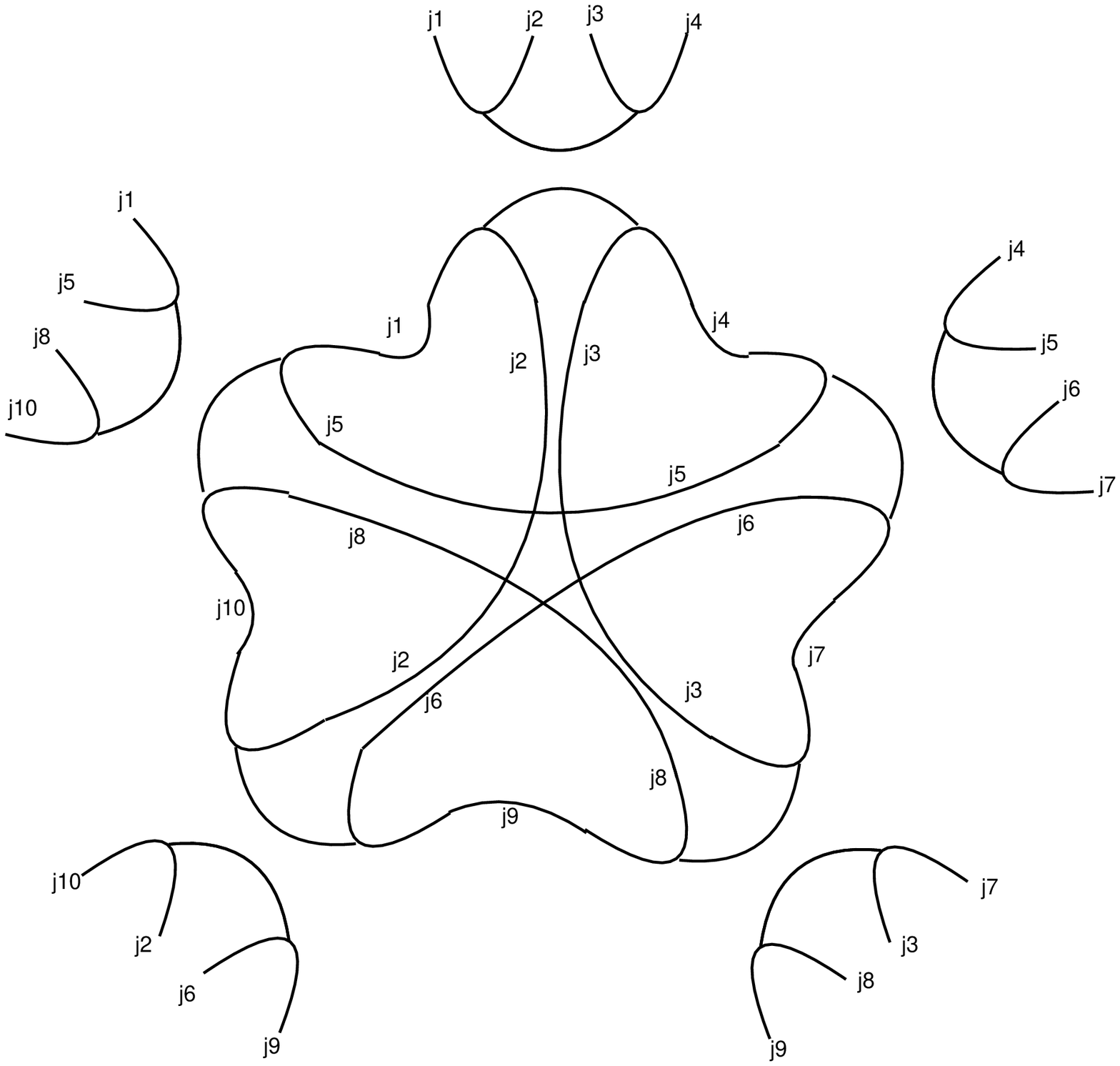}

{\normalsize Figure 6 - Schematic representation of the Barrett-Crane amplitude for a 4-simplex}
\end{center}
so that we can write down explicitly the state sum
for a manifold consisting of a single 4-simplex as:
\be
Z_{BC}=\sum_{\{j_{f}\},\{k_{e'}\}}\prod_{f}\Delta_{j_{f}}\prod_{e'}\frac{B_{k_{e'1}k_{e'2}k_{e'3}k_{e'4}}^{j_{e'1}j_{e'2}j_{e'3}j_{e'4}}}{\Delta_{j_{e'1}}\Delta_{j_{e'2}}\Delta_{j_{e'3}}\Delta_{j_{e'4}}}\prod_{v}\mathcal{B}_{BC}\;\left(\prod_{\tilde{e}}D\right)
\ee
where it is understood that there is only one vertex, $\mathcal{B}_{BC}$ is the Barrett-Crane amplitude for a 
4-simplex, and the notation $e'i$ means that we are referring to the i-th face (in some given ordering) of the 
tetrahedron $e'$, which is on the boundary of the 4-simplex, or equivalently to the i-th 2-simplex of the four which are
 incident to the dual edge (1-simplex) $e'$ of the spin foam (dual 2-complex), which is open, i.e. not ending on any 
other 4-simplex. Also the D-functions for the exposed edges are constrained to be in the simple representation.

\section{Gluing 4-simplices and the state sum for a general manifold with boundary} \label{sec:BC} 
Now consider the problem of gluing two 4-simplices together along a
common tetrahedron, say, 1234.

The most natural way to do it, having already the state sum for a single 4-simplex, so for the simplest manifold with boundary, 
is to consider the two 4-simplices separately, so considering the
common tetrahedron in the interior twice, and glue 
them together along it. So we are considering the state sum for a single 4-simplex as the basic and unique building 
block for constructing more complex state sums for more complex manifolds. 

The gluing is done by multiplying the two single partition functions, and
imposing that the values of the spins and of the projections (the $k_{e'i}$'s) of the
common tetrahedron are of course the same in the two partition
functions (this comes from the integration over the group elements assigned to the exposed edges that are being glued and become part of the interior, ans thus have to be integrated out).

Everything in the state sum is unaffected by the gluing, except for the
common tetrahedron, which now is in the interior of the manifold.
In this naive sense we could say that this way of gluing is local,
because it depends only on the parameters of the 
common tetrahedron, i.e. it should be determined only by the two
boundary terms which are associated with it when it is 
considered as part of the two different 4-simplices that are being glued.

What exactly happens for the amplitude of this interior tetrahedron
is:
\bea
\sum_{\{m\}}\frac{B_{m_{1}m_{2}m_{3}m_{4}}^{J_{1}J_{2}J_{3}J_{4}}}{\Delta_{J_{1}}\Delta_{J_{2}}\Delta_{J_{3}}\Delta_{J_{4}}}\frac{B_{m_{1}m_{2}m_{3}m_{4}}^{J_{1}J_{2}J_{3}J_{4}}}{\Delta_{J_{1}}\Delta_{J_{2}}\Delta_{J_{3}}\Delta_{J_{4}}}=\sum_{\{m\},I,L}\frac{C_{m_{1}m_{2}m_{3}m_{4}}^{J_{1}J_{2}J_{3}J_{4}I}C_{m_{1}m_{2}m_{3}m_{4}}^{J_{1}J_{2}J_{3}J_{4}L}}{\left(\Delta_{J_{1}}\Delta_{J_{2}}\Delta_{J_{3}}\Delta_{J_{4}}\right)^{2}}\nonumber
\\ = \sum_{I,L}\frac{\sqrt{\Delta_{I}\Delta_{L}}C_{m_{1}m_{2}m}^{J_{1}J_{2}I}C_{m_{3}m_{4}m}^{J_{1}J_{2}I}C_{m_{1}m_{2}n}^{J_{1}J_{2}L}C_{m_{3}m_{4}n}^{J_{3}J_{4}L}}{\left(\Delta_{J_{1}}\Delta_{J_{2}}\Delta_{J_{3}}\Delta_{J_{4}}\right)^{2}}
\nonumber \\ = \sum_{I,L}\frac{\Delta_{I}\delta_{IL}\delta_{mn}}{\left(\Delta_{J_{1}}\Delta_{J_{2}}\Delta_{J_{3}}\Delta_{J_{4}}\right)^{2}} = \sum_{I}\frac{\Delta_{I}}{\left(\Delta_{J_{1}}\Delta_{J_{2}}\Delta_{J_{3}}\Delta_{J_{4}}\right)^{2}}
\eea
where we have used the orthogonality between the intertwiners, and $I$ labels the interior edge (tetrahedron).

We see that the result of the gluing is the insertion of an
amplitude for the tetrahedra (dual edges) in the interior of the
triangulated manifold, and of course the disappearing of the boundary terms $B$ since the tetrahedron is not anymore 
part of the boundary of the new manifold (see figure 7).
\begin{center}

\includegraphics[width=5.5cm]{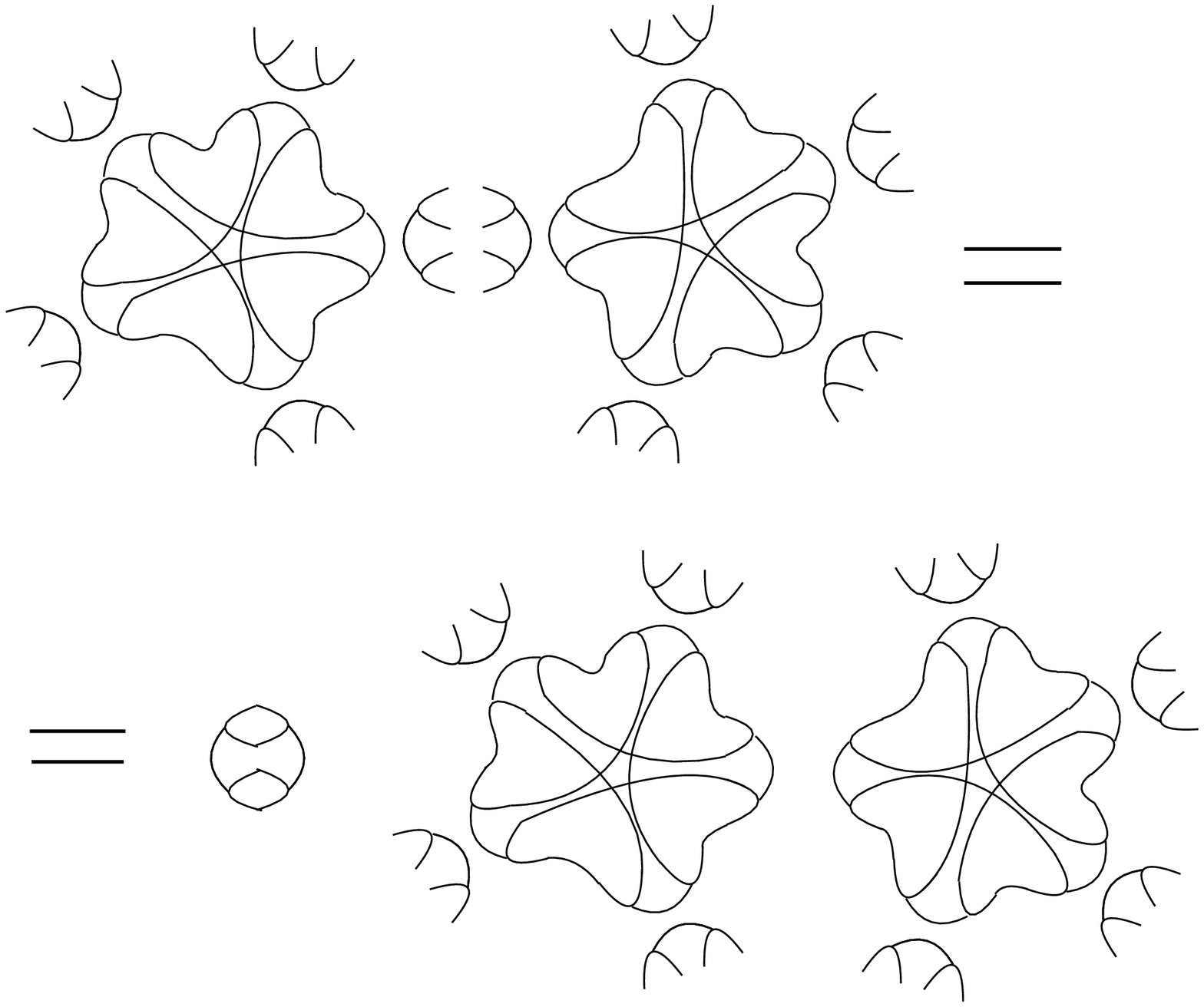}

{\normalsize Figure 7 - The gluing of two 4-simplices along a common tetrahedron}
\end{center}
We can now write down explicitly the state sum for a manifold with
boundary which is then constructed out of an arbitrary number of 4-simplices, and
has some tetrahedra on the boundary and some in the interior:
\be
Z_{BC}=\sum_{\{j_{f}\},\{k_{e'}\},\{J_{e}\}}\prod_{f}\Delta_{j_{f}}\prod_{e'}\frac{B_{k_{e'1}k_{e'2}k_{e'3}k_{e'4}}^{j_{e'1}j_{e'2}j_{e'3}j_{e'4}}}{\Delta_{j_{e'1}}\Delta_{j_{e'2}}\Delta_{j_{e'3}}\Delta_{j_{e'4}}}\prod_{e}\frac{\Delta_{J_{e}}}{\left(\Delta_{j_{e1}}\Delta_{j_{e2}}\Delta_{j_{e3}}\Delta_{j_{e4}}\right)^{2}}\prod_{v}\mathcal{B}_{BC}
\;\left( \prod_{\tilde{e}}D\right)\label{eq:final}
\ee
where the $\{e'\}$ and the $\{e\}$ are the sets of boundary and interior edges of the spin foam, respectively, while the $\tilde{e}$ are the remaining exposed edges.

It is important to note that the number of parameters which determine the gluing and that in the end characterize the 
tetrahedron in the interior of the manifold is five (4 labels for the faces and one for the tetrahedron itself), which is 
precisely the number of parameters necessary in order to determine  a first quantized geometry of a tetrahedron 
\cite{BaezBarr}. 

Moreover, the partition function with which we ended, apart from the boundary terms, is the one obtained in 
\cite{P-R}, studying a generalized matrix model, and shown to be
finite at all orders in the sum over the 
representations \cite{P-R,Per}.
 
We see that our procedure of inserting the Barrett-Crane constraints at the level of the representations starting from 
a discretized BF theory has led us exactly to the Barrett-Crane model for Euclidean quantum gravity, with a precise 
prescription for the amplitudes of the faces and, more important, the edges of the spin foam, coming directly and 
naturally from the gluing of the different 4-simplices, and also with sensible boundary terms.   

Several comments are opportune at this point.

\begin{itemize}
\item the gluing procedure used above is consistent with the formalism developed for general spin foams \cite{Baez,Baez2}, saying that when we glue two manifolds $\mathcal{M}$ and $\mathcal{M}'$ along a common 
boundary, the partition functions associated to them and to the composed manifold satisfy 
$Z(\mathcal{M})Z(\mathcal{M}')=Z(\mathcal{M}\mathcal{M}')$, as is easy to verify; see \cite{Baez,Baez2}
 for more details;
\item in the original Barrett-Crane paper \cite{BC} two different ways of considering the tetrahedra in 
the interior of the manifold were mentioned: one is to consider them separatelyas part of two different 4-simplices, and then label them
 with two Barrett-Crane intertwiners; the other one is to consider them each as a autonomous element of the 
triangulation, and so label them with only one simple representation of SO(4). Consequently there should have been 
two different models. Our result suggest that the two models are in fact the same one, because if we consider an 
interior tetrahedron as belonging to two different 4-simplices, then it is separately on the boundary of the two and 
should be assigned a boundary term as in (~\ref{eq:final}) for each of the 4-simplices. The gluing then will give us a 
term for an interior edge, as in (~\ref{eq:final}). If on the other hand we consider the whole manifold and a tetrahedron in it 
as an independent element, we should assign to it the term for an interior edge as given again in (~\ref{eq:final}); 
consequently the state sum will be the same in both cases at the end;
\item if we had imposed on our representation functions invariance under the left action of SO(3) (instead of the right 
one we used in the above) and then to the edge amplitude the invariance under the   
left action of SO(4) (again, instead of the right one used before), we would have ended 
up with exactly the same result, i.e. the same spin foam model with the same amplitude for the simplices of different 
dimensionality. There are two other possible cases. We could have imposed the invariance under the right action of 
SO(3) and the left action of SO(4), so having:
\bea
\lefteqn{A_{e}=\int_{SO(4)}dg_{e}\int_{SO(3)}dh_{1}\int_{SO(3)}dh_{2}\int_{SO(3)}dh_{3}\int_{SO(3)}dh_{4}\int_{SO(4)}dg'_{e}\,}
\nonumber \\ && D_{k_{e1}m_{e1}}^{J_{1}^{e}}(g'_{e}g_{e}h_{1})D_{k_{e2}m_{e2}}^{J_{2}^{e}}(g'_{e}g_{e}h_{2})D_{k_{e3}m_{e3}}^{J_{3}^{e}}(g'_{e}g_{e}h_{3})D_{k_{e4}m_{e4}}^{J_{4}^{e}}(g'_{e}g_{e}h_{4})\;\;\;\;\;\;\;\;
\eea 
but in this case we would have ended up with a trivial state sum model with amplitude one for each 4-simplex, 
because the multiplication of all the terms for the tetrahedra leads to a product of the norms of the unit vectors $w$, 
which are by definition one, as could be verified performing the integrals and carrying out the same steps as above. 
Otherwise we could have imposed the invariance under the left action of SO(3) and then, on the whole amplitude, the
 right invariance under SO(4):
\bea
\lefteqn{A_{e}=\int_{SO(4)}dg_{e}\int_{SO(3)}dh_{1}\int_{SO(3)}dh_{2}\int_{SO(3)}dh_{3}\int_{SO(3)}dh_{4}\int_{SO(4)}dg'_{e}\,}
\nonumber \\ && D_{k_{e1}m_{e1}}^{J_{1}^{e}}(h_{1}g_{e}g'_{e})D_{k_{e2}m_{e2}}^{J_{2}^{e}}(h_{2}g_{e}g'_{e})D_{k_{e3}m_{e3}}^{J_{3}^{e}}(h_{3}g_{e}g'_{e})D_{k_{e4}m_{e4}}^{J_{4}^{e}}(h_{4}g_{e}g'_{e})\;\;\;\;\;\;\;
\eea
obtaining, after the integrations:
\be 
A_{e}\,=\,\sum_{\Lambda}\frac{\Delta_{\Lambda}}{\sqrt{\Delta_{j_{1}}\Delta_{j_{2}}\Delta_{j_{3}}\Delta_{j_{4}}}}w^{j_{1}}_{k_{1}}w^{j_{2}}_{k_{2}}w^{j_{3}}_{k_{3}}w^{j_{4}}_{k_{4}}\,C_{m_{1}m_{2}m_{3}m_{4}}^{j_{1}j_{2}j_{3}j_{4}}\ee
and for a 4-simplex:
\be
Z=\sum_{\{j_{f}\}\{j_{e'}\}\\\{k\}}\prod_{f}\Delta_{j_{f}}\prod_{e'}\frac{\Delta_{j_{e'}}}{\sqrt{\Delta_{j_{e'1}}\Delta_{j_{e'2}}\Delta_{j_{e'3}}\Delta_{j_{e'4}}}}w^{j_{e'1}}_{k_{e'1}}w^{j_{e'2}}_{k_{e'2}}w^{j_{e'3}}_{k_{e'3}}w^{j_{e'4}}_{k_{e'4}}\;\left\{15j\right\}_{v}
\ee
where the 15j-symbol is an ordinary 15-j symbol constructed from a product of five $C$ functions. The whole 
partition function for this model (gluing different 4-simplices as above) is then:
\bea  
Z=\sum_{\{j_{f}\}\{j_{e'}\}\{k\}}\prod_{f}\Delta_{j_{f}}\prod_{e'}\frac{\Delta_{j_{e'}}}{\sqrt{\Delta_{j_{e'1}}\Delta_{j_{e'2}}\Delta_{j_{e'3}}\Delta_{j_{e'4}}}}w^{j_{e'1}}_{k_{e'1}}w^{j_{e'2}}_{k_{e'2}}w^{j_{e'3}}_{k_{e'3}}w^{j_{e'4}}_{k_{e'4}}\times \nonumber \\ \times \prod_{e}\frac{\Delta_{j_{e}}^{2}}{\Delta_{j_{e1}}\Delta_{j_{e2}}\Delta_{j_{e3}}\Delta_{j_{e4}}}\prod_{v}\left\{15j\right\}_{v}.
\eea

This looks like the case A in \cite{DP-F-K-R}, but with different amplitudes for the 2 and 3 dimensional simplices, 
and of course with additional boundary terms.

Note, however, that in this case the amplitudes for the interior
 tetrahedra are not really coming from the gluing, which
 gives just a multiplication of pre-existent factors, without any new contribution, so we could say that this alternative 
model somehow makes the gluing more trivial, because of more trivial boundary terms. In fact we could 
absorb all the boundary terms for a 4-simplex except the vectors $w$ in the vertex amplitude, as a rescaling, and 
then, after the gluing, the final state sum would not have any amplitude ($A_{e}=1$) for the interior tetrahedra. 

In addition, our result suggests that the correct and complete way of deriving a state sum  that implements the 
Barrett-Crane constraints from a generalized matrix model is like in \cite{P-R}, i.e. imposing the constraints on the 
representations only in the interaction term of the field over a group manifold, because this derivation leads to the 
correct edge amplitudes coming from the gluing, and these amplitudes are not present in \cite{DP-F-K-R}; 
\item regarding the regularization issue, it seems that even starting from a discretized action in which the sum over the 
representations is not convergent, we end up with a state sum which
is finite at all orders, according to the results of
\cite{P-R,Per}. Anyway another way to regularize completely the state sum model, making it finite at all orders, is to use a quantum group at a root of unity so that the 
sum over the representations is automatically finite due to the finiteness of the number of representations of any such quantum group; in this case we have only to replace the elements of the state sum coming from the recoupling 
theory of SO(4) (intertwiners and 15j-symbols) with the corresponding objects for the quantum deformation of it;     
\item the semiclassical limit of this state sum can be studied with the same methods as \cite{BarrWill}, leading to the 
similar results;
\item the structure of the state sum and the form of the boundary terms is a very close analogue of that discovered in 
\cite{CCM1,CCM2} for SU(2) topological field theories, like the Crane-Yetter model in 4 dimensions, in any 
number of dimensions, the difference being the group used, of course, and the absence of any constraints on the 
representations so that the topological invariance is maintained;
\item in these works, and also in ours, the boundary conditions are chosen so that the connection is fixed on the 
boundary; if another boundary condition is chosen, for example if we fix the B field to be constant on the boundary, or we choose a mixed situation where we fix part of the connection and part of the B field on the boundary, then we have to add another term in the action, and consequently the state sum will be different; an analysis of these problems was carried on in \cite{O'L} for the 3-dimensional case (Turaev-Viro model).
\end{itemize}

\section{Generalization to the case of arbitrary number of dimensions}
It is quite straighforward is to generalize our procedure and results
to an arbitrary number of dimensions of spacetime. 
Much of what we need in order to do it is already at our disposal. It was shown in \cite{F-K-P} that it is possible to 
consider gravity as a constrained BF theory in any dimension (incidentally, the same was proposed for supergravity 
\cite{Eza,L-S,L-S2}), and also the concept of simple spin networks was generalized to any dimensions in 
\cite{F-K}, with the representations of SO(D) (Spin(D)) required to be invariant under a general transformation of 
SO(D-1), so that they are realized as harmonic functions over the homogeneus space 
$SO(D)/SO(D-1)\simeq S^{D-1}$, and so that the spin network itself can be thought as a kind of Feynmann 
diagram for spacetime. 

Also the construction of a complete hierarchy of discrete topological field theories in every dimension of spacetime 
performed in \cite{CCM2}, with a structure similar to that one we propose for the Barrett-Crane model, represents 
an additional motivation for doing this. 

What is missing for applying our procedure in this general case is just a discretization of a BF theory with general 
gauge group $SO(D)$, where we cannot make use of any decomposition of the algebra in terms of the $SU(2)$ one.

Anyway, we can guess that the structure of the discretized partition function found in section ~\ref{sec:BF} for the 
group SO(4) in 4 dimensions of spacetime, using the splitting of this
group into a product of two SU(2) groups, is indeed the 
general form of a discretized partition function for BF theory for any compact group in any dimension (interestingly, it 
is analogous to that found in \cite{Kawa} for the case of SU(2) in 4 dimensions, and in \cite{KN} for SU(2) in 3 
dimensions).

So we are saying that it is reasonable to think that the discretized  partition function for BF theory in an arbitrary 
D-dimensional (Euclidean) (triangulated) spacetime for any compact group $G$ and in particular for SO(D) is:
\be
Z_{BF}\,=\,\left( \prod_{e}\int_{G}dg_{e}\right)\prod_{f}\sum_{J_{f}}\,\Delta_{J_{f}}\,\chi_{J_{f}}\left(\prod_{e'\in\partial f}g_{e'}\right)
\ee
where $e'$ is a dual link on the boundary of the dual plaquette $f$ (face of the spin foam) associated with a triangle 
$t$ in the triangulation, $e$ indicates the set of dual links, and the character is in the representation $J_{f}$ of the 
group $G$.

A way to justify this heuristically (for a more rigorous discretization leading to the same result, see \cite{FK})  is the 
following. 

We start from a discretized action like the one we used before (of course the same remarks concerning the approximation
 used apply also here), 

\be S_{BF}\,=\,\sum_{t}\,B(t)\,F(t) \ee having again approximated the $B$ field with a distributional field with values 
only on the (D-2)-simplices t of the original triangulation, and with: $e^{iF(t)}=\prod_{e'\in\partial f}g_{e'}$. 
With this action the partition function for the theory becomes:
\bea
\lefteqn{Z_{BF}\,=\,\int_{G}\mathcal{D}A\int_{G}\mathcal{D}B(t)\,e^{i\sum_{t}B(t)F(t)}} \nonumber \\ 
&=&\,\int_{G}\mathcal{D}A\int_{G}\mathcal{D}B(t)\prod_{t}e^{i B(t)F(t)}\,=\,\int_{G}\mathcal{D}A\,\prod_{t}\,\delta\left( e^{iF(t)}\right) \nonumber \\
&=&\, \prod_{e}\int_{G}dg_{e}\prod_{f}\delta\left( \prod_{e'\in\partial f}g_{e'}\right) 
\eea
with the notation as above, and having replaced the product over the (D-2)-simplices with a product over the faces of the 
dual triangulation (plaquette), that is possible because they are in 1-1 correspondence.

Now we can use the decomposition of the delta function of a group
element into a sum of characters, obtaining:
\be
Z_{BF}\,=\,\left( \prod_{e}\int_{G}dg_{e}\right)\prod_{f}\sum_{J_{f}}\,\Delta_{J_{f}}\,\chi_{J_{f}}\left(\prod_{e'\in\partial f}g_{e'}\right)
\ee
i.e. the partition function we were trying to derive.

From now on we can proceed as for SO(4) in 4 dimensions. Consider $G=SO(D)$, and the $J$'s as the highest 
weight labelling the representations of that group.

We can decompose the characters into a product of D-functions, and rearrange the sums and products in the 
partition function to obtain:
\be 
Z_{BF}\,=\,\sum_{\{J_{f}\}, \{k\},\{m\}}\prod_{f}\Delta_{f}\prod_{e}\,A_{e}\left(\prod_{\tilde{e}}D\right) \ee
where
\be
A_{e}\,=\,\int_{SO(D)}dg_{e}D_{k_{e1}m_{e1}}^{J_{e1}}(g_{e})...D_{k_{eD}m_{eD}}^{J_{eD}}(g_{e}) \ee
where $ei$ labels the i-th of the D faces incident on the edge $e$.

Now we can apply our procedure and insert here the Barrett-Crane constraints:
\bea
\lefteqn{A_{e}\,=\,\int_{SO(D)}dg_{e}\int_{SO(D-1)}dh_{1}...\int_{SO(D-1)}dh_{D}\int_{SO(D)}dg'_{e}}
\nonumber \\ &&D_{k_{e1}m_{e1}}^{J_{e1}}(g_{e}h_{1}g'_{e})...D_{k_{eD}m_{eD}}^{J_{eD}}(g_{e}h_{D}g'_{e}). \eea

Performing the integrals, and carrying on the same steps as in
 Sections ~\ref{sec:1BC} and~\ref{sec:BC}, leads to the
 analogue of the formula (~\ref{eq:final}) in higher dimensions (or
 alternatively to the analogue of case A in \cite{DP-F-K-R}):
\bea
\lefteqn{Z_{BC}^{D}\,=\,\sum_{\{j_{f}\},\{J_{e}\},\{J_{e'}\},\{k_{e'}\},\{K_{e'}\}}\,\prod_{f}
 \Delta_{j_{f}}\prod_{e'}\sqrt{\Delta_{J_{e'1}}...\Delta_{J_{e'(D-3)}}}}\nonumber\\ &&\frac{C^{j_{e'1}j_{e'2}J_{e'1}}_{k_{e'1}k_{e'2}K_{e'1}}C^{j_{e'3}J_{e'2}J_{e'1}}_{k_{e'3}K_{e'2}K_{e'1}}...C^{j_{e'(D-2)}J_{e'(D-4)}J_{e'(D-3)}}_{k_{e'(D-2)}K_{e'(D-4)}K_{e'(D-3)}}C^{j_{e'(D-1)}j_{e'D}J_{e'(D-3)}}_{k_{e'(D-1)}k_{e'D}K_{e'(D-3)}}}{\Delta_{j_{e'1}}...\Delta_{j_{e'D}}} \nonumber \\ && \prod_{e}\frac{\Delta_{J_{e1}}...\Delta_{J_{e(D-3)}}}{\left( \Delta_{j_{e1}}...\Delta_{j_{eD}}\right)^{2}}\,\prod_{v}\mathcal{B}_{BC}^{D}\;\left(\prod_{\tilde{e}}D\right).
\eea
There are D faces (corresponding to (D-2)-dimensional simplices) incident on each edge (corresponding to 
(D-1)-dimensional simplices) $e$ ((D-1)-simplex in the interior) or $e'$ ((D-1)-simplex on the boundary).  
There are D+1 edges for each vertex (corresponding to a D-dimensional simplex), and consequently $D(D+1)/2$ 
faces for each D-simplex. Each edge is labelled by a set of (D-3) $J$'s. $\mathcal{B}_{BC}^{D}$ is the higher 
dimensional analogue of the Barrett-Crane amplitude, i.e. (the SO(D) analogue of) the $\frac{3}{2}(D+1)(D-2)J$-symbol constructed out of the $D(D+1)/2$ labels of the faces and the $(D-3)(D+1)$ labels of the edges.

Again, this result is a very close analogue of the state sum for a topological field theory in general dimension, obtained 
in \cite{CCM2}.

Of course, everywhere we are summing over only simple representations of SO(D), i.e. representations of SO(D) that
 are of class 1 with respect to the subgroup SO(D-1) \cite{V-K}.
 
\section{Conclusions}
We conclude with a summary of our procedure and results, and with some comments on possible directions of future 
work.

In this paper we have proposed a way to derive the Barrett-Crane spin foam model for Euclidean quantum gravity in 
4 dimensions, starting from a discretization of a SO(4) BF theory (SO(4) is the local symmetry group in the Euclidean 
case). The trick is to discretize the (unconstrained) classical BF theory, and then impose the constraints that lead to 
the gravity theory at the quantum level, which means at the level of the representations of the gauge group by which 
we label the elements of the spin foam. In this way, we argue, it is possible to circumvent the difficulties in discretizing 
and then quantizing directly the Plebanski action for gravity, which is the classical counterpart of the Barrett-Crane 
model.  
The result we end with is exactly the Barrett-Crane spin foam model with a precise prescription of the form the the
 Barrett-Crane state sum should have, in the general case of an arbitrary manifold with boundary. In particular we 
derived the amplitude for the edges of the spin foam, from a clear and natural procedure of gluing different 4-simplices
 together along a common tetrahedron. The fact that our result coincide with that derived in \cite{P-R} from a 
generalized matrix model, and shown to be necessary to make the sum over colorings finite, seems to 
reinforce our proposal. Moreover our results and the state sum we obtain can be easily generalized to higher 
dimensions.

We can now say something about possible ways to improve and develop our results, and to apply them to some 
physical problem.

Of course a very important thing to do would be to discretize the Plebanski action, so to start with a constrained BF 
theory at the classical level, and quantize it in the same way we did, to see if we really end up with the same result, as
 we expect.

Another thing that should be studied more and that should be quite straighforward is a derivation, along the same lines, 
of a Barrett-Crane state sum for a different choice of boundary conditions, for example with the B field kept fixed on 
the boundary instead of the connection. To obtain this we have only to discretize the additional boundary action in 
the BF theory and apply again our procedure.

Having at hand both kinds of state sum, a careful analysis of the more appropriate boundary conditions in the case of a
 spacetime representing a black hole should be carried out, and then in principle it should be possible (apart from the 
technical problems of the calculations) to use the derived state sum model for a calculation of black hole entropy, along
 the lines of \cite{BHETV} for the 3-dimensional case. 

Of course, before trying to apply the state sum model to any physical problem, another thing is necessary, that is the 
implementation in a precise way of a sum over triangulations, or over spin foams (2-complexes), which is necessary 
to restore an infinite number of degrees of freedom to the theory. Only after this is done, we can start to consider this 
model as a concrete proposal for a quantum theory of gravity, so regarding it as an approach to a complete 
quantization of Einstein theory, leading to the possibility of studying physical aspects of gravitational field in a sensible 
way. A very promising approach to this problem is represented by the generalized matrix models proposed in 
\cite{DP-F-K-R,P-R}, and generalized to any kind of spin foam (also not related to gravity) in \cite{R-R1,R-R2}, but more study is necessary to settle it down definitely.

Still to understand and develop is a Lorentzian (causal) version of these models; some work in this direction was 
carried on in \cite{BC2}, and recently important results were found in \cite{P-R2,P-R3}. Our procedure should be in 
principle (apart from all the technical difficulties) applicable also to the case in which the local symmetry group is the 
Lorentz group, and this possibility will be investigated in the future. A very different but related solution to the problem 
of implementing causality in these models was proposed and developed in \cite{M-S,M,M-S2}. 

Also still to understand is the question of the semiclassical limit of the model. In fact, apart from the connection with 
discrete gravity obtained in \cite{BarrWill}, a great deal of work is still necessary to understand how a classical 
background metric can emerge from a theory of this kind, and if it is possible, as has to be possible, to develop a 
picture of perturbations of this background from this model, proving
that it contains gravitons as is necessary for 
any satisfactory theory of quantum gravity. 

\section*{Acknowledgements}
This work was supported by PPARC. Both authors are
grateful to the Theory Division for hospitality at CERN, where part of this work was
done. D.O. thanks J. W. Barrett for useful discussions and
comments. D.O. also acknowledges ICRA for partial financial support,
and St.Edmund's college for a travel grant.

\end{document}